\documentclass[twocolumn,showpacs,preprintnumbers]{revtex4}%
\usepackage{amssymb}
\usepackage{amsmath}
\usepackage{graphicx}
\usepackage{epstopdf}
\usepackage{dcolumn}
\usepackage{bm}
\usepackage{amsfonts}%
\providecommand{\U}[1]{\protect\rule{.1in}{.1in}}
\providecommand{\U}[1]{\protect\rule{.1in}{.1in}}
\begin{document}
\title{Sensing Dispersive and Dissipative Forces by an Optomechanical Cavity}
\author{Oren Suchoi}
\affiliation{Department of Electrical Engineering, Technion, Haifa 32000 Israel}
\author{Eyal Buks}
\affiliation{Department of Electrical Engineering, Technion, Haifa 32000 Israel}
\date{\today }

\begin{abstract}
We experimentally study an optomechanical cavity that is formed between a
mechanical resonator, which serves as a movable mirror, and a stationary
on-fiber dielectric mirror. A significant change in the behavior of the system
is observed when the distance between the fiber's tip and the mechanical
resonator is made smaller than about 1 $%
\operatorname{\mu m}%
$. The observed effects are attributed to the combined influence of Casimir
force, Coulomb interaction due to trapped charges, and optomechanical
coupling. The comparison between experimental results and theory yields a
partial agreement.

\end{abstract}
\maketitle





The study of the interaction between a mechanical resonator and nearby bodies
is of great importance for the fields of micro electromechanical systems and
scanning probe microscopy. For sufficiently short distances the interaction is
expected to be dominated by the Casimir force,
\cite{Lamoreaux_5,Mohideen_4549,Chan_211801}, which originates from the
dependence of the ground state energy of the electromagnetic field upon
boundary conditions
\cite{Casimir_793,Lifshitz_73,Milonni_Quantum_Vacuum,Bordag_1,Lamoreaux_850,Lamoreaux_201}%
. For larger distances, however, other mechanisms such as Coulomb interaction
between trapped charges and their image charges \cite{Denk_2171} and local
variations in the work function \cite{Burnham_144} commonly dominate the interaction.%

\begin{figure}
[ptb]
\begin{center}
\includegraphics[
height=2.5949in,
width=3.4537in
]%
{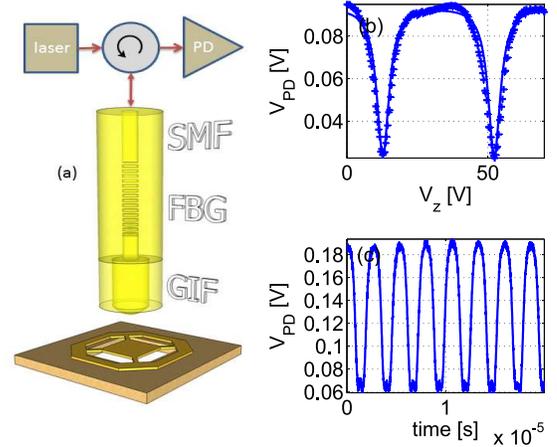}%
\caption{The experimental setup. (a) The fiber probe is composed of a FBG
mirror and a GIF lens having a tip that has been blown into a dome shape. A
tunable laser excites the cavity and the off reflected optical power is
measured using a photodetector (PD). (b) The PD voltage $V_{\mathrm{PD}}$ vs.
the voltage $V_{\mathrm{z}}$ that is applied to the piezoelectric motor
controlling the distance between the dome and the trampoline. The crosses
represent experimental results, which have been obtained with injected optical
power of $P_{\mathrm{L}}=0.90\operatorname{mW}$ at wavelength $\lambda
=1545.525\operatorname{nm}$. The solid line represents the theoretically
predicted voltage, which is obtained from the calculated reflection
probability $R_{\mathrm{C}}=1-I\left(  x\right)  /\beta_{\mathrm{F}}$, where
$I\left(  x\right)  $ is given by Eq. (\ref{I(x)}), with the parameters
$\beta_{+}=0.3$ and $\beta_{-}=0.15$. (c) The measured PD voltage
$V_{\mathrm{PD}}$ vs. time in the region of SEO with injected optical power of
$P_{\mathrm{L}}=1.77\operatorname{mW}$ at the same wavelength.}%
\label{Fig setup}%
\end{center}
\end{figure}

In this study we investigate the effect of interaction between nearby bodies
on the dynamics of an optomechanical cavity \cite{Braginsky_653,
Hane_179,Gigan_67,Metzger_1002,Kippenberg_1172,Favero_104101,Marquardt2009}.
In our setup the optomechanical cavity is formed between two mirrors, a
stationary fiber Bragg grating (FBG) mirror and a movable mirror made of
aluminum in the shape of a trampoline supported by 4 beams [see Fig
\ref{Fig setup}(a)]. The tip of the fiber is blown into a dome shape.
Piezoelectric motors are employed for positioning the center of the dome above
the center of the trampoline and for controlling the distance $d$ between
them. The observed response of the optomechanical cavity in the range
$d\lesssim1%
\operatorname{\mu m}%
$ exhibits rich dynamics resulting from the interplay between back-reaction
optomechanical effects and the nonlinear coupling between the interacting
bodies. In general, such coupling may result in both a static force due to
dispersive interaction, and a friction force due to dissipative (or retarded)
interaction \cite{Volokitin_345}. Contrary to some other previously employed
techniques, in which only the static force can be measured, we find that the
observed response of the optomechanical cavity allows the extraction of both
static and friction forces \cite{Stipe_096801,Dorofeyev_2402,Gotsmann_2597}.

A photo-lithography process is used to pattern a $200%
\operatorname{nm}%
$ thick aluminum on a high resistivity silicon wafer, into a mechanical
resonator in the shape of a $100\times100%
\operatorname{\mu m}%
^{2}$ trampoline \cite{Zaitsev_046605} [see Fig. \ref{Fig setup}(a)]. Details
of the fabrication process can be found elsewhere \cite{Suchoi_033818}.
Measurements are performed at a temperature of $77%
\operatorname{K}%
$ and a pressure well below $5\times10^{-5}%
\operatorname{mbar}%
$. A graded index fiber (GIF) of $0.47$ pitch is spliced to the end of the
single mode fiber (SMF), and its tip is blown into a dome shape of radius
$R=90%
\operatorname{\mu m}%
$. A cryogenic piezoelectric three-axis positioning system having
sub-nanometer resolution is employed for manipulating the position of the
optical fiber. A tunable laser operating near the Bragg wavelength
$\lambda_{\mathrm{B}}=1545.7%
\operatorname{nm}%
$ of the FBG together with an external attenuator are employed to excite the
optical cavity. The optical power reflected off the cavity is measured by a
photodetector (PD), which is connected to both a spectrum analyzer and to an
oscilloscope. Two neighboring optical cavity resonances are seen in panel (b)
of Fig. \ref{Fig setup}, in which the reflected optical power is plotted as a
function of the voltage $V_{\mathrm{z}}$ that is applied to the piezoelectric
motor, which is employed for controlling the vertical distance between the
dome and the trampoline. A time trace in the region of self-excited
oscillation (SEO) is shown in panel (c).

The technique of resonance sensing allows measuring both static and friction
forces acting on the trampoline mirror. The linear response of the decoupled
fundamental mechanical mode of the trampoline is characterized by a complex
angular frequency $\Upsilon_{0}=\omega_{\mathrm{m}}-i\gamma_{\mathrm{m}}$,
where $\omega_{\mathrm{m}}=2\pi\times381.9%
\operatorname{kHz}%
$ is the intrinsic angular resonance frequency of the mode and $\gamma
_{\mathrm{m}}=1.5%
\operatorname{Hz}%
$ is its intrinsic damping rate. In general, interaction between the
mechanical mode and a given ancilla system may give rise to an external force
acting on the mechanical resonator. For a fixed mechanical displacement $x$
the force is characterized by its static value, which is denoted by
$F_{\mathrm{s}}\left(  x\right)  $. For simplicity, it is assumed that the
time evolution of the ancilla system is governed by a first order equation of
motion, which is characterized by a decay rate $\gamma_{\mathrm{s}}$. To
lowest nonvanishing order in the coupling strength between the mechanical
resonator and the ancilla system the effect of the interaction effectively modifies the
value of the complex angular resonance frequency, which becomes $\Upsilon
_{\mathrm{eff}}=\Upsilon_{0}+\Upsilon_{\mathrm{s}}$, where the contribution
due to back-reaction $\Upsilon_{\mathrm{s}}$ is give by \cite{Zaitsev_1589}%
\begin{equation}
\Upsilon_{\mathrm{s}}=\frac{\gamma_{\mathrm{s}}F_{\mathrm{s}}^{\prime}\left(
x_{\mathrm{f}}\right)  }{2m\omega_{\mathrm{m}}}\frac{1}{\gamma_{\mathrm{s}%
}+i\omega_{\mathrm{m}}}\;, \label{Upsilon_ba}%
\end{equation}
where $m$ is the effective mass of the fundamental mechanical mode,
$x_{\mathrm{f}}$ is the displacement of the mechanical resonator corresponding
to a stationary solution of the equations of motion, and it is assumed that $\gamma_{\mathrm{m}}\ll
\omega_{\mathrm{m}}$.

In the discussion below, the contribution to $\Upsilon_{\mathrm{s}}$ due to
Casimir force is denoted by $\Upsilon_{\mathrm{C}}=\omega_{\mathrm{C}}%
-i\gamma_{\mathrm{C}}$, the one due to Coulomb interaction induced by trapped
charges by $\Upsilon_{\mathrm{T}}=\omega_{\mathrm{T}}-i\gamma_{\mathrm{T}}$,
and the one due to optomechanical coupling by $\Upsilon
_{\mathrm{B}}=\omega_{\mathrm{B}}-i\gamma_{\mathrm{B}}$. Both contributions
$\Upsilon_{\mathrm{C}}$ and $\Upsilon_{\mathrm{T}}$ become negligibly small
when the dome-trampoline distance $d$ is sufficiently large, whereas the
contribution of the optomechanical term $\Upsilon_{\mathrm{B}}$ can be suppressed
by reducing the laser power $P_{\mathrm{L}}$.

\begin{figure}
[ptb]
\begin{center}
\includegraphics[
height=2.8098in,
width=3.4546in
]%
{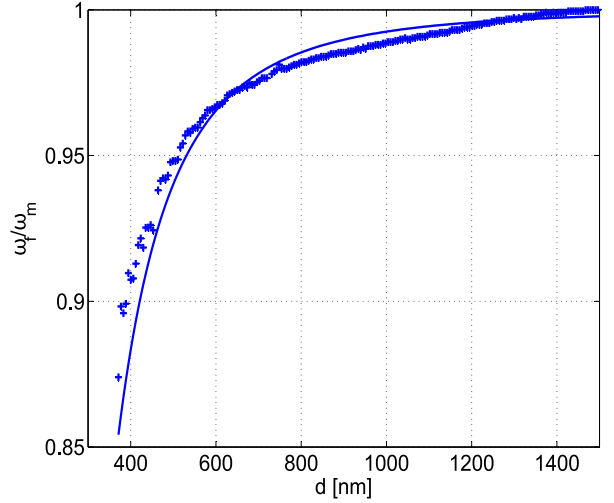}%
\caption{The normalized mechanical resonance frequency $\omega_{\mathrm{f}%
}/\omega_{\mathrm{m}}$ vs. distance $d$. The solid line represents the
theoretical prediction based on Eqs. (\ref{Upsilon_ba}), (\ref{F_C}) and
(\ref{F_T PC}), whereas experimentally measured values are labelled by crosses
. The assumed values of experimental parameters that have been used in the
calculation are $\epsilon=n_{\mathrm{GIF}}^{2}$, where $n_{\mathrm{GIF}}=1.49$
is the refractive index of the silica-made GIF, $d_{\mathrm{p}}%
=6.2\operatorname{nm}$, $R=90\operatorname{\mu m}$, $m=1.3\times
10^{-11}\operatorname{kg}$ and $q_{\mathrm{T}}=1.1\times10^{-14}%
\operatorname{C}$.}%
\label{Fig Freq vs. d}%
\end{center}
\end{figure}

The normalized mechanical resonance frequency $\omega_{\mathrm{f}}%
/\omega_{\mathrm{m}}$ is plotted in Fig. \ref{Fig Freq vs. d} as a function of
the dome-trampoline distance $d$. The measured values, which are obtained with
laser power of $P_{\mathrm{L}}=0.39%
\operatorname{mW}%
$ and laser wavelength of $1545.05%
\operatorname{nm}%
$ (the laser is tuned away from the Bragg band of high reflectivity), are labelled
by crosses. For these laser parameters optomechanical back-reaction effects
are experimentally found to be negligibly small, allowing thus isolating the
combined contributions of $\Upsilon_{\mathrm{C}}$ and $\Upsilon_{\mathrm{T}}$.
To theoretically estimate these contributions, both the Casimir force
$F_{\mathrm{C}}\left(  d\right)  $ and the Coulomb force due to trapped
charges $F_{\mathrm{T}}\left(  d\right)  $ are evaluated below as a function
of the distance $d$ between the dome and the trampoline.%

The Casimir force per unit area $P_{\mathrm{PP}}\left(  d\right)  $ between a
metal plate having plasma frequency $\omega_{\mathrm{p}}$ and a dielectric
plate having a relative dielectric constant $\epsilon$ separated by a vacuum
gap of width $d$ can be evaluated using the Lifshitz formula
\cite{Lifshitz_73,Milonni_Quantum_Vacuum,Bordag_1,Chen_020101,buhmann2007dispersion}%
\begin{equation}
P_{\mathrm{PP}}\left(  d\right)  =-\frac{3\hbar c\left(  \epsilon-1\right)
}{32\pi^{2}d^{4}}I_{\mathrm{L}}\left(  \frac{d}{d_{\mathrm{p}}},\epsilon
-1\right)  \;, \label{P_PP}%
\end{equation}
where $d_{\mathrm{p}}=c/2\omega_{\mathrm{p}}$, the function $I_{\mathrm{L}}$
is given by%
\begin{align}
I_{\mathrm{L}}\left(  D,y\right)   &  =\int_{1}^{\infty}\mathrm{d}p\;\int
_{0}^{\infty}\mathrm{d}x\;\frac{x^{3}}{3p^{2}y}\nonumber\\
&  \times\left(  \frac{1}{\zeta_{\mathrm{s},1}\zeta_{\mathrm{s},2}e^{x}%
-1}+\frac{1}{\zeta_{\mathrm{p},1}\zeta_{\mathrm{p},2}e^{x}-1}\right)
\;,\nonumber\\
&  \label{I_L}%
\end{align}
where%
\begin{align}
\zeta_{\mathrm{s},1}  &  =\frac{1+\sqrt{1+\left(  \frac{D}{x}\right)  ^{2}}%
}{1-\sqrt{1+\left(  \frac{D}{x}\right)  ^{2}}}\;,\\
\zeta_{\mathrm{p},1}  &  =\frac{1+\left(  \frac{pD}{x}\right)  ^{2}%
+\sqrt{1+\left(  \frac{D}{x}\right)  ^{2}}}{1+\left(  \frac{pD}{x}\right)
^{2}-\sqrt{1+\left(  \frac{D}{x}\right)  ^{2}}}\;,
\end{align}
and where%
\begin{align}
\zeta_{\mathrm{s},2}  &  =\frac{p+\sqrt{p^{2}+y}}{p-\sqrt{p^{2}+y}}\;,\\
\zeta_{\mathrm{p},2}  &  =\frac{\left(  1+y\right)  p+\sqrt{p^{2}+y}}{\left(
1+y\right)  p-\sqrt{p^{2}+y}}\;.
\end{align}
Note that $I_{\mathrm{L}}\left(  D,y\right)  \rightarrow1$ in the limit
$D\rightarrow\infty$ and $y\rightarrow0$. In Eq. (\ref{P_PP}) the effect of
absorption in the dielectric material has been disregarded and the correction
due to finite temperature has been neglected. These approximations are
expected to be valid provided that $\hbar c/\Delta_{\mathrm{D}}\ll d\ll\hbar
c/k_{\mathrm{B}}T$ , where $\Delta_{\mathrm{D}}$ is the energy gap of the
dielectric material. For the parameters of the current experiment the validity
condition reads $22%
\operatorname{nm}%
\ll d\ll30%
\operatorname{\mu m}%
$.

When the distance $d$ between the metallic trampoline and the dielectric dome
is much smaller than the radius of the dome $R$ the mutual force, which is
labeled by $F_{\mathrm{C}}\left(  d\right)  $, can be evaluated using the
Derjaguin approximation \cite{Derjaguin_819} [see Eq. (\ref{P_PP})]%
\begin{equation}
F_{\mathrm{C}}\left(  d\right)  =2\pi R\int_{d}^{\infty}\mathrm{d}%
z\;P_{\mathrm{PP}}\left(  z\right)  \;. \label{F_C}%
\end{equation}

Finite metal conductivity may give rise to a friction force associated with
the Casimir interaction \cite{Volokitin_345}. The effect of Casimir friction
on the mechanical resonator can be characterized by a damping rate, which is
denoted by $\gamma_{\mathrm{C}}$. For the parameters of our device the
theoretical expression given in Ref. \cite{Volokitin_345} yields a value
$\gamma_{\mathrm{C}}/\omega_{\mathrm{m}}\simeq10^{-12}$, which is about 7
orders of magnitude smaller than the intrinsic mechanical quality factor, and
thus the Casimir friction is not expected to play any significant role in the
current experiment \cite{Stipe_096801,Dorofeyev_2402,Gotsmann_2597}.

Coulomb interaction between trapped charges and their images may give rise to
an additional force acting on the mechanical resonator \cite{Stipe_096801}. In general the force depends on the
unknown distribution of trapped charges. In what follows, it is assumed that
the force can be expressed in terms of an effective total trapped charge
$q_{\mathrm{T}}$ as \cite{Marchi_538}%
\begin{equation}
F_{\mathrm{T}}\left(  d\right)  =\frac{q_{\mathrm{T}}^{2}}{4\pi\epsilon
_{0}\left(  2d\right)  ^{2}}\;, \label{F_T PC}%
\end{equation}
where $\epsilon_{0}$ is the permittivity constant. Note that for the case
where all trapped charges are located on the surface of the dome at the point
closest to the trampoline Eq. (\ref{F_T PC}) becomes exact provided that
polarizability can be disregarded.

In general, trapped charges can give rise to both, a shift in the effective
value of the angular frequency of the mechanical resonator, which is denoted
by $\omega_{\mathrm{T}}$, and to an added damping rate, which is denoted by
$\gamma_{\mathrm{T}}$. The added damping rate can be evaluated by calculating
the damping power generated by dissipative currents on the surface of the
metal due to relative motion of trapped charges
\cite{Stowe_2785,Chumak_085407}. The ratio $\gamma_{\mathrm{T}}/\left\vert
\omega_{\mathrm{T}}\right\vert $ is found to be roughly given by
$\gamma_{\mathrm{T}}/\left\vert \omega_{\mathrm{T}}\right\vert \simeq
4d\omega_{\mathrm{m}}/\lambda_{\mathrm{D}}\sigma$, where $\lambda_{\mathrm{D}%
}$ is the Debye length ($\simeq1.7\times10^{-10}%
\operatorname{m}%
$ for aluminum) and where $\sigma$ is the conductivity ($\simeq3.0\times
10^{18}%
\operatorname{Hz}%
$ for aluminum at $77%
\operatorname{K}%
$). For the entire range of values of the distance $d$ that has been explored
in the current experiment $\gamma_{\mathrm{T}}/\left\vert \omega_{\mathrm{T}%
}\right\vert <3\times10^{-8}$, and thus the added damping due to trapped
charges is expected to be negligibly small. Moreover, retardation in the
redistribution of charges on the surface of the metal due to mechanical motion
can be safely disregarded since $\sigma\gg\omega_{\mathrm{m}}$.

The complex frequency shift $\Upsilon_{\mathrm{C}}+\Upsilon_{\mathrm{T}}$
induced by the combined effect of Casimir interaction and trapped charges can
be evaluated using Eq. (\ref{Upsilon_ba}). As was discussed above, the
imaginary part of both $\Upsilon_{\mathrm{C}}$ and $\Upsilon_{\mathrm{T}}$ can
be safely disregarded. The fixed point value of the displacement of the
mechanical resonator, which is denoted by $x_{\mathrm{f}}$, is found by
solving the force balance equation $m\omega_{\mathrm{m}}^{2}x=F_{\mathrm{C}%
}\left(  d-x\right)  +F_{\mathrm{T}}\left(  d-x\right)  $. The normalized
measured (crosses) and calculated (solid line) values of the mechanical
angular resonance frequency $\omega_{\mathrm{f}}$ are plotted in Fig.
\ref{Fig Freq vs. d} as a function of the distance $d$. The assumed values of
experimental parameters that have been used in the calculation are listed in
the caption of Fig. \ref{Fig Freq vs. d}.

As can be seen from Fig. \ref{Fig Freq vs. d}, the smallest measured value of
$\omega_{\mathrm{f}}/\omega_{\mathrm{m}}$ prior to pull-in is about $0.87$. As
was previously pointed out in Ref. \cite{Buks_220}, pull-in due to thermal
activation (which is estimated to be much more efficient than quantum
tunneling for the parameters of the current experiment) is theoretically
expected to occur at significantly smaller values of the ratio $\omega
_{\mathrm{f}}/\omega_{\mathrm{m}}$.

In the above-discussed measurements back-reaction effects originating from
coupling between the optical cavity and the mechanical resonator have been
suppressed by keeping the injected optical power $P_{\mathrm{L}}$ at a
relatively low level. Such effects, however, can significantly modify the
dynamics at higher values of $P_{\mathrm{L}}$. In general, the effect of
radiation pressure typically governs the optomechanical coupling when the
finesse of the optical cavity is sufficiently high
\cite{Kippenberg_et_al_05,Rokhsari2005,
Arcizet2006,Gigan_67,Cooling_Kleckner06, Kippenberg_1172}, whereas, bolometric
effects can contribute to the optomechanical coupling when optical absorption
by the vibrating mirror is significant \cite{Metzger_1002,
Jourdan_et_al_08,Marino&Marin2011PRE, Metzger_133903, Restrepo_860,
Liberato_et_al_10,Marquardt_103901, Paternostro_et_al_06,Yuvaraj_430}, and
when the thermal relaxation rate is comparable to the mechanical resonance
frequency \cite{Aubin_1018, Marquardt_103901, Paternostro_et_al_06,
Liberato_et_al_10_PRA}. Bolometric optomechanical coupling
\cite{Metzger_133903, Metzger_1002, Aubin_1018,
Jourdan_et_al_08,Zaitsev_046605,Zaitsev_1589} may result in many intriguing
phenomena such as mode cooling and SEO
\cite{Hane_179,Kim_1454225,Aubin_1018,Carmon_223902,Marquardt_103901,Corbitt_021802,Carmon_123901,Metzger_133903,Bagheri_726,Rugar1989,
Arcizet2006a, Forstner2012,Weig2013}.

In the device under study in the current experiment the dominant underlying
mechanism responsible for the optomechanical coupling is optical absorption by
the suspended mirror \cite{Zaitsev_046605}. Such absorption gives rise to
heating, which in-turn causes thermal deformation of the suspended structure
due to mismatch between thermal expansion coefficients of the suspended mirror
made of aluminum and the supporting silicon substrate \cite{Yuvaraj_430}.
Thermal deformation \cite{Metzger_133903} gives rise to a thermal force, which
is expressed as $m\theta T_{\mathrm{R}}$, where $\theta$ is assumed to be a
constant, and where $T_{\mathrm{R}}=T-T_{0}$ is the offset between the
temperature of the suspended mirror $T$ and the temperature of the supporting
substrate $T_{0}$. In the static limit the force, which is denoted for this
case by $F_{\mathrm{B}}$, can be evaluated by simultaneously solving the force
balance equation $\omega_{\mathrm{m}}^{2}x=\theta T_{\mathrm{R}}$, where $x$
denotes the mechanical displacement, and the thermal balance equation
$Q=\gamma_{\mathrm{H}}T_{\mathrm{R}}$, where $Q$ is the heating power divided
by the thermal heat capacity of the trampoline and where $\gamma_{\mathrm{H}}$
is the thermal decay rate.

Optical interference in the cavity gives rise to displacement dependence of
the term $Q$ , which is given by $Q=\eta P_{\mathrm{L}}I\left(  x\right)  $,
where $\eta$ is the heating coefficient due to optical absorption and where
$P_{\mathrm{L}}I\left(  x\right)  $ is the intra-cavity optical power incident
on the suspended mirror. The function $I\left(  x\right)  $ depends on the
properties of the optical cavity. The finesse of the optical cavity is limited
by loss mechanisms that give rise to optical energy leaking out of the cavity.
The main escape routes are through the on-fiber static reflector, through
absorption by the metallic mirror, and through radiation. The corresponding
transmission probabilities are respectively denoted by $\mathcal{T}%
_{\mathrm{B}}$, $\mathcal{T}_{\mathrm{A}}$ and $\mathcal{T}_{\mathrm{R}}$. In
terms of these parameters, the function $I\left(  x\right)  $ is given by
\cite{Zaitsev_046605}%
\begin{equation}
I\left(  x\right)  =\frac{\beta_{\mathrm{F}}\left(  1-\frac{\beta_{-}^{2}%
}{\beta_{+}^{2}}\right)  \beta_{+}^{2}}{1-\cos\frac{4\pi x_{\mathrm{D}}%
}{\lambda}+\beta_{+}^{2}}\;, \label{I(x)}%
\end{equation}
where $x_{\mathrm{D}}=x-x_{\mathrm{R}}$ is the displacement of the mirror
relative to a point $x_{\mathrm{R}}$, at which the energy stored in the
optical cavity in steady state obtains a local maximum, $\beta_{\pm}%
^{2}=\left(  \mathcal{T}_{\mathrm{B}}\pm\mathcal{T}_{\mathrm{A}}\pm
\mathcal{T}_{\mathrm{R}}\right)  ^{2}/8$ and where $\beta_{\mathrm{F}}$ is the
cavity finesse. The reflection probability is given in steady state by $R_{\mathrm{C}}=1-I\left(
x\right)  /\beta_{\mathrm{F}}$ \cite{Yurke_5054,Zaitsev_046605}.%

\begin{figure}
[ptb]
\begin{center}
\includegraphics[
height=3.0026in,
width=3.4537in
]%
{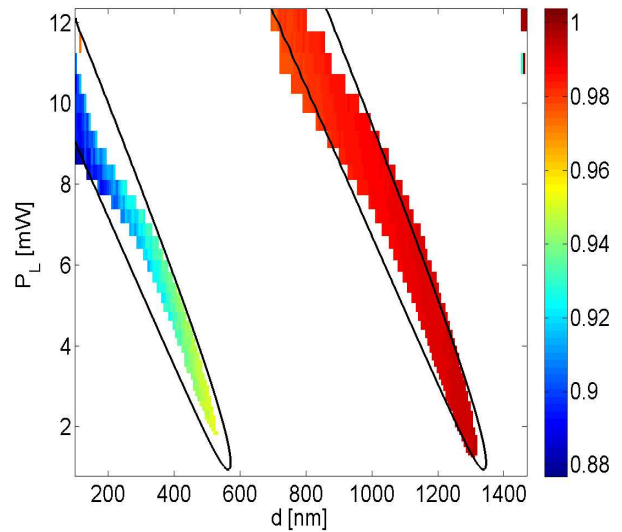}%
\caption{The measured normalized value of SEO frequency $\omega_{\mathrm{SEO}%
}/\omega_{\mathrm{m}}$ vs. dome-trampoline distance $d$ and laser power
$P_{\mathrm{L}}$. The black solid lines represent the theoretically calculated
bifurcation lines, along which the effective damping rate vanishes. The
assumed values of experimental parameters that have been used in the
calculation are $\beta_{\mathrm{F}}=3.0$ and $\omega_{\mathrm{m}}^{2}%
\gamma_{\mathrm{H}}\lambda/\theta\eta=3.3\operatorname{mW}$ (see also the
captions of Figs. \ref{Fig setup} and \ref{Fig Freq vs. d}).}%
\label{Fig SEO_Freq}%
\end{center}
\end{figure}

With sufficiently high laser power the system can be driven into SEO. The
color-coded plot seen in Fig. \ref{Fig SEO_Freq} exhibits the measured
normalized value of SEO frequency $\omega_{\mathrm{SEO}}/\omega_{\mathrm{m}}$
vs. dome-trampoline distance $d$ and laser power $P_{\mathrm{L}}$. No SEO is
observed in the white regions. The two colored regions in Fig.
\ref{Fig SEO_Freq} represent SEO occurring near two optical resonances (OR).
The one seen on the left is the first OR of the cavity, which occurs with the
smallest value of $d$, and the one seen on the right is the second one. As can
be seen from Fig. \ref{Fig SEO_Freq}, the SEO frequency $\omega_{\mathrm{SEO}%
}$ measured near the first OR is significantly smaller. Moreover, the lowest
input laser power value for which SEO occurs near the first OR is
significantly higher (the value is $1.45$ times larger than the value
corresponding to the second OR).

The black solid lines in Fig. \ref{Fig SEO_Freq} represent the theoretically
calculated bifurcation lines, which are found from solving the equation
$\gamma_{\mathrm{m}}+\gamma_{\mathrm{B}}=0$, where $\gamma_{\mathrm{B}%
}=-\operatorname*{imag}\Upsilon_{\mathrm{B}}$, and $\Upsilon_{\mathrm{B}}$,
which is given by Eq. (\ref{Upsilon_ba}), is calculated for the case of the
above-discussed bolometric coupling between the mechanical resonator and the
optical cavity. In spite of the fact that the contribution of both Casimir and
Coulomb interactions to the effective mechanical damping rate is theoretically
expected to be negligibly small, the experimental results clearly indicate
that the damping rate is significantly larger near the first OR, as can be seen from the significantly higher observed value of laser power threshold . Further
study is needed in order to identify the underlying mechanism responsible for
this contactless friction that is observed at relatively short distances.

In summary, sensitive detection of both dispersive and dissipative forces is
demonstrated using an optomechanical cavity. The combined effect of Casimir
force, Coulomb interaction due to trapped charges and bolometric
optomechanical coupling on the mechanical resonator is theoretically
estimated. Partial agreement is found in the comparison between theory and
experimental findings.

This work was supported by the Israel Science Foundation, the Binational
Science Foundation, the Security Research Foundation at Technion and the
Russell Berrie Nanotechnology Institute.

\bibliographystyle{apsrev}
\bibliography{C:/Users/eyal/software/swp40/TCITeX/BibTeX/bib/Eyal_bib/Eyal_Bib}

\end{document}